\documentclass[aps,prf,nofootinbib,twocolumn,showkeys,showpacs]{revtex4}

\usepackage{amsmath}
\usepackage[utf8]{inputenc}
\usepackage{caption}
\usepackage{graphicx}
\usepackage{multirow}
\usepackage{xcolor}
\usepackage{color}
\usepackage{aas_macros}
\usepackage{ulem}

\begin{document}
\title{Deep learning for  multimessenger core-collapse supernova detection}
\author{M. L\'{o}pez Portilla\footnote{This research was conducted during my studies at Maastricht University.}$^{1,2}$ , I. Di Palma$^{3a,b}$, M. Drago$^{3a,4a,b}$, P. Cerd\'{a}-Dur\'{a}n$^5$, F. Ricci$^{3a,b}$}
\affiliation{$^1$ Institute for Gravitational and Subatomic Physics (GRASP), 
Department of Physics, Utrecht University, 
Princetonplein 1, 3584 CC Utrecht, The Netherlands}
\affiliation{$^{2}$ Nikhef, Science Park 105, 1098 XG Amsterdam, The Netherlands}
\affiliation{$^{3a}$ Universit\`a di Roma  {\it{La Sapienza}}, I-00185 Roma, Italy}
\affiliation{$^{3b}$ INFN, Sezione di Roma, I-00185 Roma, Italy}
\affiliation{$^{4a}$ Gran Sasso Science Institute (GSSI), I-67100 L'Aquila, Italy}
\affiliation{$^{4b}$ INFN, Laboratori Nazionali del Gran Sasso, I-67100 Assergi, Italy}
\affiliation{$^5$ Departamento de Astronom\'ia y Astrof\'isica, Universitat de Val\`encia, Dr. Moliner 50, 46100, Burjassot (Valencia), Spain}

\begin{abstract}
 The detection of gravitational waves from core-collapse supernova (CCSN) explosions is a challenging task, yet to be achieved, in which it is key the connection between multiple messengers, including neutrinos and electromagnetic signals. 
 In this work, we present a method for detecting these kind of signals based on machine learning techniques. We tested its robustness by injecting signals in the real noise data taken by the Advanced LIGO-Virgo network during the second observation run, O2. We trained a newly developed Mini-Inception Resnet neural network using time-frequency images corresponding to injections of simulated phenomenological signals, which mimic the waveforms obtained in 3D numerical simulations of CCSNe. With this algorithm we were able to identify signals from both our phenomenological template bank and from actual numerical 3D simulations of CCSNe.
 We computed the detection efficiency versus the source distance, obtaining that, for signal to noise ratio  higher than 15, the detection efficiency is 70 \% at a false alarm rate lower than 5\%.
We notice also that, in the case of O2 run, it would have been possible to detect signals emitted at 1 kpc of distance, whilst lowering down the efficiency to 60\%, the event distance reaches values up to 14 kpc.
 
\end{abstract}
\keywords{Gravitational waves, Core collapse supernovae, Multimessenger Astrophysics.}
\maketitle


\section{Introduction}

The recent discovery of gravitational waves and high-energy cosmic neutrinos, marked the beginning  of a new era of the multimessenger astronomy. These new messengers, along with electromagnetic radiation and cosmic rays, give new insights into the most extreme energetic cosmic events. Among them supernovae explosion is one of the challenging targets of this new astronomical approach.   

The advanced generation of gravitational wave detectors has proved their capacity of detecting gravitational waves (GWs) from astrophysical processes.
While gravitational waves have been detected from mergers of binary black holes and binary neutron stars, core collapse supernovae (CCSN), have not been detected yet and they still represent a puzzle to solve.
We had confirmation of the basic CCSN theory through the detection of MeV neutrinos from the SN1987A \cite{SN1987A}: the collapse of a massive star's core is driven by the release of gravitational energy and the vast majority of this energy is realised in neutrinos.
However, the details of the mechanism of the explosion are still an open question and the astronomical community is trying to disentangle the supernova explosion mechanism with intense studies.

Massive stars $(M>8 M_\odot)$ spend most of their lives burning hydrogen into helium, which settles in the core and, when temperatures increase sufficiently, burns into heavier nuclei until iron is reached.
The resulting iron core is supported by electron degeneracy pressure. When the core reaches the Chandrasekhar mass, pressure cannot support its own weight anymore and collapses. The collapse of the inner core is stopped abruptly when nuclear saturation density is reached at the center and the inner core bounces back. 
Shortly after the core bounce neutrino emission carries away energy from the post-shock layer.
In the present models of the neutrino driven supernovae explosions,  the intense hydrodynamic mass motion has to play a significant role in the energy transfer by the neutrino flux to the the medium behind the stagnating core-bounce shock, reviving the outward shock motion and thus initiates the SN explosion. Due to the weak coupling of neutrinos in the region of this energy deposition, in the  hydrodynamic models of the explosions  a large variety of  physical ingredients are needed \citep{Janka,Janka2}. 
This so called neutrino driven mechanism \cite{Bethe}, is the dominant theory to explain CCSN explosions in slowly rotating progenitors.  
Observationally only $\sim1\%$ of the events shows signatures of fast rotation (broad-lined type Ic SNe \citep{Li:2011b} or long GRBs \citep{Chapman:2007}), therefore neutrino-driven
explosions are likely the most common type of CCSN and we will focus this work on those. 

In a supernova explosion, GWs are generated in the inner core of the source, so that this messenger  carries direct information of   the inner mechanism. 
The feasibility of this scenario will be supported by the joint observation of neutrino and gravitational wave emission from CCSN, by assessing the correlation between neutrino emission and collapsed core motion.
Although the phenomenon is among of the most energetic in the universe, the amplitude of the gravitational wave impinging on a detector on the Earth is extremely faint. For a CCSN in the center of the Milky way, a rare event, we could  expect amplitudes of the metric tensor perturbations ranging between $10^{-21}- 10^{-23}$. 
To increase the detection probability we should increase the volume of the universe to be explored and this can be achieved both by decreasing the detector noise and using better performing statistical algorithms. 

The impossibility of using template-matching techniques in this case, due to the complexity and stochasticity of the waveform,
makes it necessary to find new ways to improve the detection statistics.
Current efforts to search for gravitational waves from CCSN include targeted searches for observed nearby SNe \citep{SNTargeted2016,Abbott:2019pxc} and all-sky generic searches for bursts \citep{Abbott:2016ezn,Abbott:2019prv}.
For the latter  two independent pipelines are used: coherent Waveburst (cWB) \citep{Klimenko:2015ypf} and omicron-LIB (oLiB) \citep{Lynch:2015}, while  BayesWave \citep{Cornish:2014kda} is a followup of cWB GW candidate events. These searches use algorithms based on excess power to identify signals buried in the detector's noise without taking advantage  of any specific feature of CCSN waveform.

In \cite{Rome} it has been proposed the use of machine learning techniques to take advantage of the peculiarities of the CCSN GW signal with the goal of
increasing our detection capability with respect to current methods. In  particular, the focus was on the  monotonic raise of the GW signal in the time-frequency plane due to the g-mode excitation,
which is the dominant feature present in the GW spectrum. A similar approach has been followed recently by \cite{Chan:2019,Cavaglia:2020,Iess:2020} and in general there has been an increasing interest in the GW community 
for the use of machine learning methods \cite[see][for a review]{Cuoco:2020}.

{In this paper we follow a similar approach as in \cite{Rome}, labeled in the whole paper as \textit{previous work.}}
The main differences are \begin{itemize} 
\item[-] the use of a more sophisticated convolutional neural network (CNN);
\item[-] the injection of simulated CCSN signals in real noise of the three advanced detectors of the LIGO-Virgo network, as measured during August 2017 (the previous work only considered Gaussian noise);
\item[-] the improvement of the phenomenological templates used during the training of the CNN network to better match results from numerical simulations.\end{itemize}

This paper is structured as follows.
In section II we describe our newly improved phenomenological waveform templates that are used to train the CNN networks presented in section III. In section IV we describe the detector noise data used for the injections.
Section V is devoted to  the procedure of the training of the CNN network and its behaviour. In section VI we report the results, showing the detection performance in terms of Signal to Noise Ratio and event distance. Results are discussed in section VII  and then we conclude.


\section{Waveforms}
\label{sec:waveforms}

To implement our search method,  we have pursued  an approach similar to the \textit{previous work} \cite{Rome}. We consider a parametric phenomenological waveform designed to match the most common features observed in the numerical models of CCSN.  
We focus our attention on the g-modes excitation, the most common feature of all models developed so far to describe the CCSN phenomena, responsible for the bulk of the GW signal in the post-bounce evolution of the proto-neutron star. The aim of our phenomenological template is to mimic the raising arch observed in core-collapse simulations. To this end we will consider a damped harmonic oscillator with a random forcing, in which the frequency varies with time.
The phenomenological templates used in this work differ with respect to the ones in \cite{Rome} in two aspects: we use a new and more flexible parametrization for the frequency evolution and we use the distance as a parameter.
The phenomenological templates are calibrated to mimic the features in the numerical simulations for non-rotating progenitor
stars by \cite{Murphy:2009,Marek:2009,Yakunin:2010,Scheidegger:2010,Muller2012,BMueller:2013,Yakunin:2015,Kuroda,Andresen2017}, named \textit{waveform calibration set}, hereafter.

The new parametrization describes the evolution of the frequency of the g-modes, $f(t)$, as a splines interpolation to a series of discrete points, $(t_{i}, f_i)$, where $t_i$ corresponds to post-bounce times. Given the relatively simple behaviour of $f(t)$ observed in numerical simulations, it is sufficient to use three points with $t_i =(0, 1, 1.5)$~s. $f_0$, $f_1$, and $f_2$ are then three new parameters of the template.

\begin{figure}[h!]
    \centering
    \includegraphics[width=0.5\textwidth]{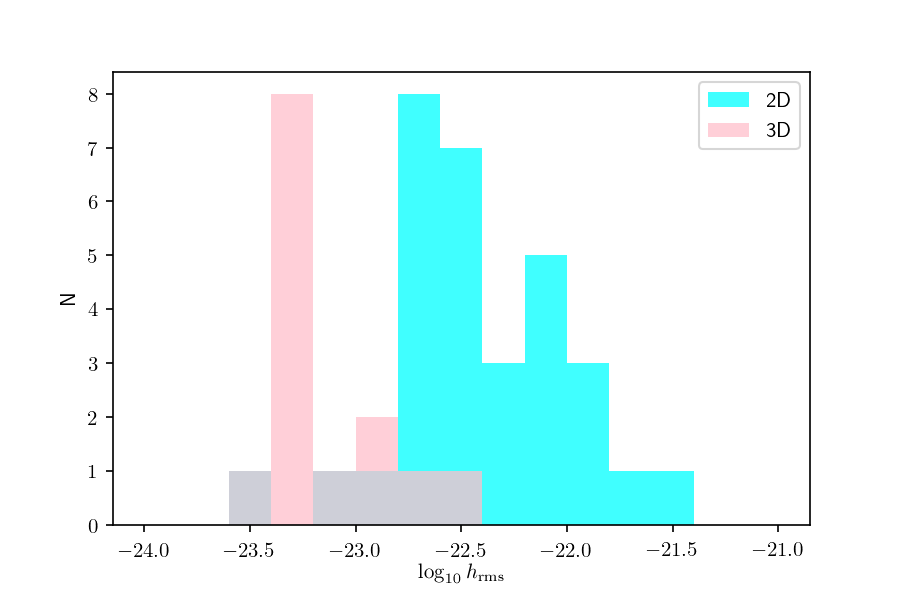}
    \captionof{figure}{Number of simulations with a given g-mode rms strain at $10$~kpc (per logarithmic interval) for 2D (blue bars) and 3D (red bars) simulations in the waveform calibration set.}
    \label{fig:RMS10kp}
\end{figure}

In \cite{Rome} the amplitude of the generated waveforms has been chosen according to the SNR. In this work we want to go one step further and use distance as a parameter for the waveform generator. In order to do that we relate the amplitude of the waveform with its distance using the data in the waveform calibration set.  First we have to measure for each simulation the typical strain of the component of the signal related to g-modes. To this aim we apply a high pass filter at
$200$~Hz, and then we use the section of the waveform containing $99\%$ of the waveform energy to compute the rms value. This procedure filters out signal power at lower frequencies related to other effects different to g-modes (standing-shock accretion
instabilities, prompt convection and large scale asymmetries due to shock propagation) that are not considered for this work. Fig.~\ref{fig:RMS10kp} shows distribution of the logarithm of the rms strain for 2D and 3D simulations at $10$~kpc.  There are significant 
differences between different simulations depending on the dimensionality. The mean and standard deviation for each distribution is  $\log_{10} h_{\rm rms, 2D} = -22.4 \pm 0.42$ and $\log_{10} h_{\rm rms, 3D} = -23.1 \pm 0.29$, for the  
2D and 3D case, respectively. Given that 3D simulations are more realistic we use this normalization to generate our phenomenological waveforms. As consequence, we end up with template amplitudes about a factor $5$ smaller than typical 2D simulations. 
In our waveform generator, the strain of each of the waveforms is scaled to have a rms strain corresponding to a random value following a normal distribution with the mean and standard deviation of our normalization, and scaled to the corresponging distance.

In synthesis, we have a waveform template that depends on a set of 8 free parameters as reported in table \ref{tab:param}.
Additionally, for any combination of those parameters we can generate multiple realisations due to the random component in the
excitation of the harmonic oscillator and on the random value of the rms strain. 
In order to be able to represent the variety of g-mode features observed in the waveform calibration set, we provide ranges covering
all the possibilities (see table~\ref{tab:param}). To this parameter space one has to add additional restrictions to ensure the monotonicity ($f_2>f_1>f_0$) and
convexity ($(f_1-f_0)/(t_1-t_0) \ge (f_2-f_1)/(t_2-t_1)$) of $f(t)$, as seen in the numerical simulations. 
We have created the waveform template bank that contains $504$ different realisations of this parameter set, for each distance, resulting of applying 
the restrictions above to the $9072$ possible combinations of the parameters
in table \ref{tab:param}.
In this way we obtain a reasonably dense covering of the parameter space.

\begin{table}
\caption{Parameter space of the phenomenological templates. The second third  and fourth columns
indicate the range (maximum and minimum, respectively) for each parameter and the spacing used in the sampling
of the parameter space. For $Q$ and $D$ we show the actual values instead. All times are post-bounce times.} 
\begin{tabular}{l | c c  c | c }
\hline
parameter & min. & max. & $\Delta$ & description\\ \hline
$t_{\rm ini}$ [s] & $0$ & $0.2$ & $0.1$&beginning of the waveform \\ 
$t_{\rm end}$ [s] & $0.2$ & $1.5$ & $0.1$ &end of the waveform\\ 
$f_0$ [Hz] & $50$ & $150$ & $50$& frequency at bounce\\ 
$f_1$ [Hz] & $1000$ & $2000$ & $500$ &frequency at $1$~s\\ 
$f_2$ [Hz] & $1500$ & $4500$ & $1000$ &frequency at $1.5$~s\\
$f_{\rm driver}$ [Hz] & $100$ & $200$ & $100$ &driver frequency\\ 
$Q$  & \multicolumn{3}{c|}{$(1, 5, 10)$} &quality factor\\ 
$D$ [kpc] & \multicolumn{3}{c|}{$(1, 2, 5, 10, 15)$}   &distance to source\\
\hline
\end{tabular}
\label{tab:param}
\end{table}



\section{Methodology}

\subsection{Challenges and milestones of Deep Learning}

The application of Deep Learning (DL) across science domains is a booming enterprise. DL algorithms have been very successful in a variety of tasks and in recent times it has emerged as a new tool in the GW field. These methods are able to perform analysis rapidly since all the intensive computation is diverted to the one-time training stage, which could make them orders of magnitude faster than conventional matched filtering technique. In addition, there are no limitations in the size of the templates bank of GW signals, and even more, it is preferable to use large data sets to cover as deep a parameter space as possible. Due to this fact they sparked the interest of several authors, who have built deep-learning algorithms to demonstrated their power on specific examples,
including CCSN \cite{Chan:2019,Cavaglia:2020,Iess:2020} among others \citep[see e.g.][]{Carillo, George_Huerta, Krastev, Cuoco:2020}.

A CNN is a specialized kind of DL algorithm to process data that has a known grid-like topology and  can learn to differentiate a variety of input types due to its ability for pattern recognition \cite{Goodfellow-et-al-2016}. In a CNN, the input is convolved with a filter, which varies according to the characteristics of the data since it can be \textit{learned} by the network. The computations are performed at each step, as the filter is slided onto the input to compute the corresponding value in the output feature map. Despite of the automatic learning of the filter, some parameters need to be tuned by hand.

The input of 2-dimensional CNN are images, which have 3 dimensions: width $w_{in}$, height $h_{in}$ and depth $d_{in}$. Assume that an image with dimensions $(w_{in}, h_{in}, d_{in})$ is convolved  with $f$ filters of size $k \times k$. The amount of pixels that the filter slides at each step is the stride $s$, while the border of zeros added has a width $p$, called convolutional padding. The result of the convolution yields the following output: 

\begin{align}\label{eq:CNN}
  \begin{pmatrix}
    w_{out}\\
    h_{out}\\
    d_{out}
  \end{pmatrix}=
  \begin{pmatrix}
    \Big[\frac{w_{in}+p-k}{s}\Big]+1\\
    \Big[\frac{h_{in}+p-k}{s}\Big]+1\\
    f
  \end{pmatrix}
\end{align}

We can also calculate the number of  parameters that we need to train for each layer (or level) as $(k \times k \times d_{in} +1)\times d_{out}$.
It is interesting to note that each layer of the CNN looks at different patterns since they can \textit{learn} different filters, depending on the information provided by the previous layers. 
Thus, these layers learn to recognize visual patterns by first extracting local features and subsequently combining them to obtain higher-level representations. 

With these ideas in mind, the \textit{previous work} provided a clear evidence  that, under relatively simplified conditions, deep CNN algorithms could  be more efficient to extract GW signals from CCSNe than the current  methodology. Therefore, the aim of this work is to improve the neural network developed in \cite{Rome}, going deeper with convolutions to increase accuracy while keeping computational complexity at a reasonable cost.

\begin{figure}[h!]
    \centering
    \includegraphics[width=7cm]{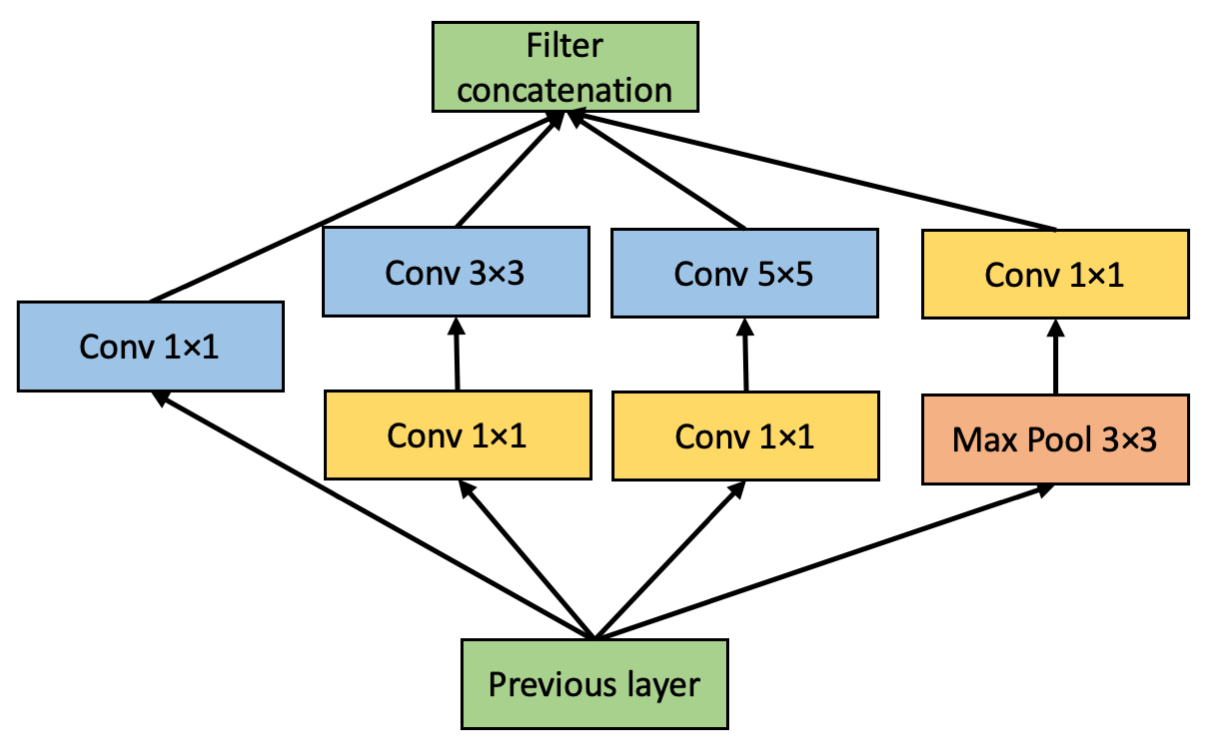}
    \captionof{figure}{Inception module with dimensionality reduction, adapted from \cite{GoogLeNet}.}
    \label{fig:googlenet}
\end{figure}

The most straightforward way of improving the performance of a deep neural network is by increasing their size, which includes the number of layers and the number of neurons per layer. Nonetheless, enlarging a network implies to train a larger amount of parameters and over-complicating the model, which increases dramatically  the computational cost. A fundamental way of solving these issues would be to move from fully connected to sparsely connected architectures, as it is discussed in \cite{GoogLeNet}. 
In this work, it is presented a sophisticated network topology construction, the so-called Inception network, that tries to approximate a sparse structure. The architecture is composed by blocks of convolutions, known as Inception modules.  The input of each block is convolved in parallel by separate CNN layers with different kernels, while the outputs of all the convolutions are stacked, as we can observe in Fig. \ref{fig:googlenet}. In such a way, a sparse network is built without the necessity of choosing a particular kernel size, but computational complexity increases drastically. 
To prevent a high computational cost the authors introduce \textit{dimensionality reduction},  i.e. $1\times1$ convolutions that reduce the depth of the output. If we convolve our input $(w_{in}, h_{in}, d_{in})$ with $f$ filter $1 \times 1$, stride $s=1$ and padding $p=0$, according to Eq. \ref{eq:CNN} the output will be $(w_{in}, h_{in}, f)$. Therefore, if $f < d_{in}$ the depth and the number of parameters will be greatly reduced. In later releases of the Inception network, the authors explore further the idea of \textit{dimensionality reduction}.

In \cite{factorization}, they explore other ways of factorizing convolutions in various settings, especially in order to increase the computational efficiency of the solution without reducing the expressiveness of the block.
Firstly, the authors examine the factorization into smaller convolutions, where they claim that $5\times5$ convolution can be \textit{factorized} into two $3\times3$ convolutions since the final output has the same dimensions. Nonetheless, the main difference between both processes are the number of parameters. A $5 \times 5$ convolution needs $(5^2 \times d_{in}+1) \times d_{out}$ parameters to train, while for two $3 \times 3$ convolutions it is necessary to train $2 \times (3^2 \times d_{in}+1) \times d_{out}$ parameters, which is less computationally expensive.
Secondly, they analyze the factorization into asymmetric convolutions, such that a convolution $c \times c \longrightarrow  c \times 1 \text{ and } 1 \times c $ convolutions. Again, the outputs of both processes have the same dimensionality but different amount of trainable parameters, $(c^2 \times d_{in}+1) \times d_{out} >  2 \times (c \times 1 \times d_{in}+1) \times d_{out}$. 
Therefore, in \cite{factorization} the authors \textit{factorize} $5 \times 5$ convolutions into $3 \times 3$, which in turn are \textit{factorized} by $3\times1$ and $1\times3$ convolutions,  to lighten the computations.

Another obstacle of deeper networks is the degradation problem, where with increasing depth, accuracy gets saturated and then degrades rapidly. In 
\cite{ResNet} this problem is approached by introducing a deep neural network, called Residual Network or ResNet.
This network is able to learn the identity function using shortcut connections that skip one or more layers, which are also known as ``skip connections''. Therefore, the network is reminded every few layers how was the input a few layers before, which can be translated in learning the identity function with a simple demonstration.  Furthermore, in \cite{ResNet} different empirical results show that the degradation problem is well addressed since accuracy gains are obtained from increasing depth.

Due to the improvements in accuracy
obtained with Inception network and Resnet, in \cite{Inception_ResNet} it was explored the combination of these two brilliant architectures, while  \textit{factorization} discussed in \cite{factorization}. As a result,  they developed, among others, an architecture called Inception-Resnet v1 which is $\sim 90$ layers depth. It was demonstrated that the introduction of residual connections lead to a dramatic improvement in the computational speed, while it was shown that Inception-Resnet algorithms were able to achieve higher accuracies with less iterations of the training phase.

Our problem is much simpler than the task performed in \cite{Inception_ResNet}, since we only need to discriminate between two classes: templates that contains a GW CCSN signal (event class) and templates that do not contain a GW CCSN signal (noise class). 
As a consequence, we have developed reduced (``mini'') versions of Inception v3, Resnet and Inception-Resnet v1, using the original building blocks of those networks, but adapting them to our needs. Since our reduced version of  Inception-Resnet v1 (Mini Inception-Resnet, hereafter) provided the best performance in all our tests, here we  only present results for this case. We describe the algorithm architecture in the following subsection.

\subsection{Architecture of Mini Inception-Resnet}

For the development of our Mini Inception-Resnet network, including the model definition, 
 the training and the validation phases, we have used the Keras frameworks \cite{cite_keras}, based on the TensorFlow backend \cite{cite_tensorflow}.
 We employ Adam optimizer \cite{Kingma:2014vow} with a learning rate $lr=0.001$ and $\epsilon=10^{-6}$ to avoid divisions by zero when computing back-propagation. The activation functions of all the convolutional layers is $relu$ activation function, $ReLU(x)= \max{(0,x)}$. We employ a batch size of $64$ because, for our particular task, it is a good trade-off between computational complexity and performance.  

Despite of facing a classification problem with two classes,  the approach used in \cite{Rome}  is to employ the categorical cross-entropy loss function with a softmax activation function in the last layer, i.e. the problem is treated as a multi-class classification problem with two classes. In this work we  simplify this approach by using a \textit{binary cross-entropy} instead and a sigmoid  activation function for the output, i.e. we address the problem as  a classification problem with a positive class (event class) and a negative class (noise class). 
It is important to note that  \textit{categorical cross-entropy} and softmax activation function are the generalizations of \textit{binary cross-entropy} and a sigmoid  activation function,  respectively.

\begin{figure}[b!th]
    \centering
    \includegraphics[width=4cm]{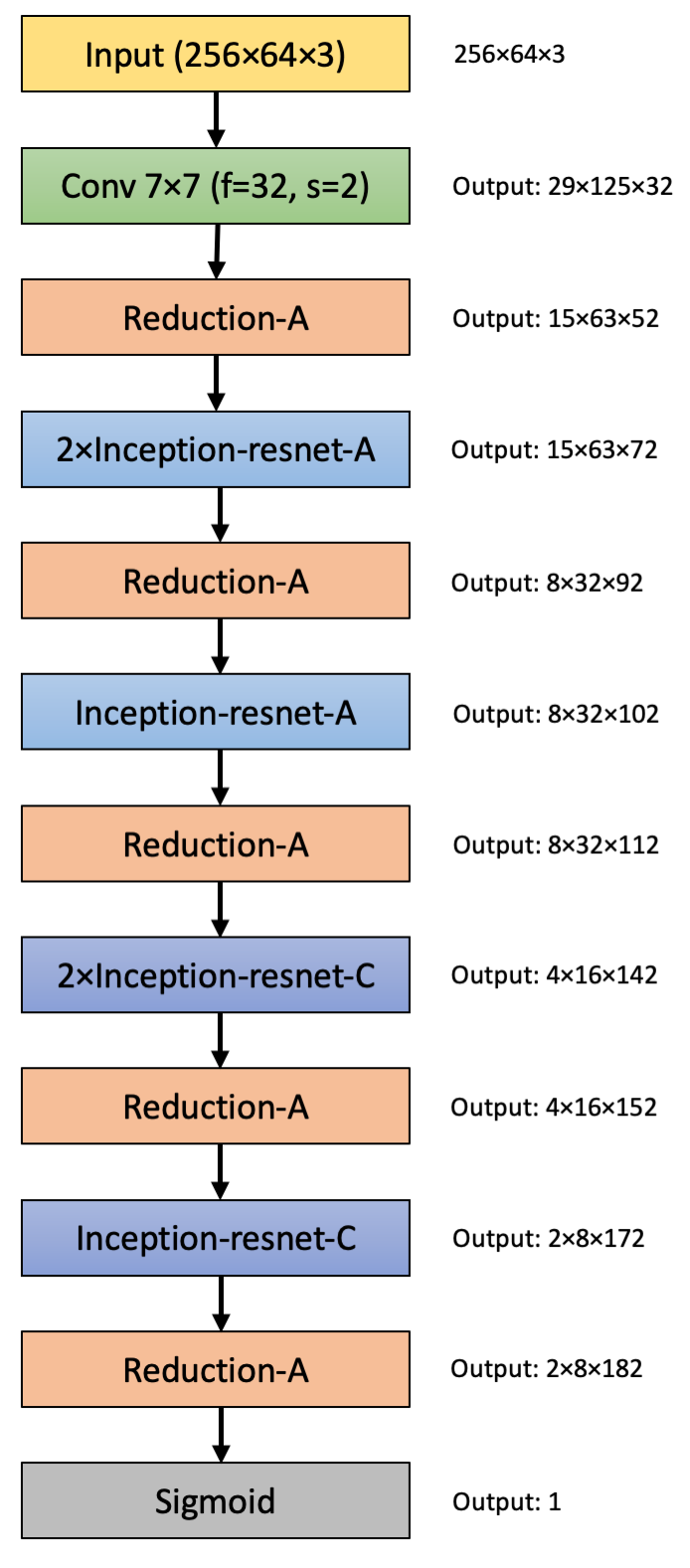}
    \captionof{figure}{The overall schema of the Mini Inception-Resnet network. For the detailed modules, please refer to Figs. \ref{fig:A}, \ref{fig:C} and \ref{fig:reductions}.}
    \label{fig:architecture}
\end{figure}

In \cite{Inception_ResNet}, the authors build 5 different types of blocks, namely Inception-ResNet-A, Inception-ResNet-B, Inception-ResNet-C, Reduction-A and Reduction-B. The modules Inception-ResNet-B and Reduction-B are the most expensive blocks, since the convolutions inside them are $1\times7$, $7\times1$ and $7\times7$. Hence,  we  discard these modules to implement the reduced version of this algorithm. At the same time, we shrink the amount of parameters of our network by interspersing Inception-Resnet modules with Reduction-A blocks (Fig. \ref{fig:architecture}).

\begin{figure}[hbt!]
    \centering
    \includegraphics[width=5.5cm]{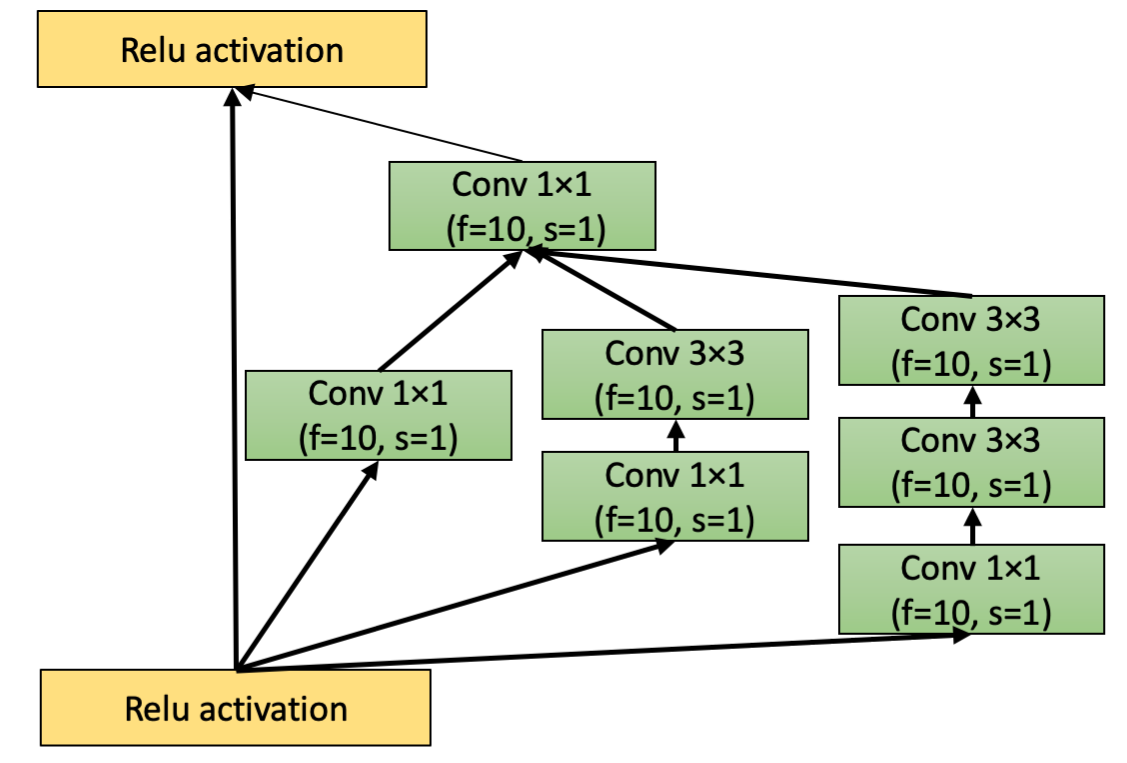}
    \captionof{figure}{The schema for Inception-ResNet-A, adapted from \cite{Inception_ResNet}.}
    \label{fig:A}
\end{figure}

The Inception-ResNet-A block (see Fig. \ref{fig:A}) is equivalent to the Inception module shown in Fig. \ref{fig:googlenet}. It is interesting to note that the Max Pooling layer is substituted by the ``shortcut connection'', and the $5\times5$ convolution is \textit{factorized} by two $3\times3$ convolution layers. 

\begin{figure}[hbt!]
    \centering
    \includegraphics[width=4.5cm]{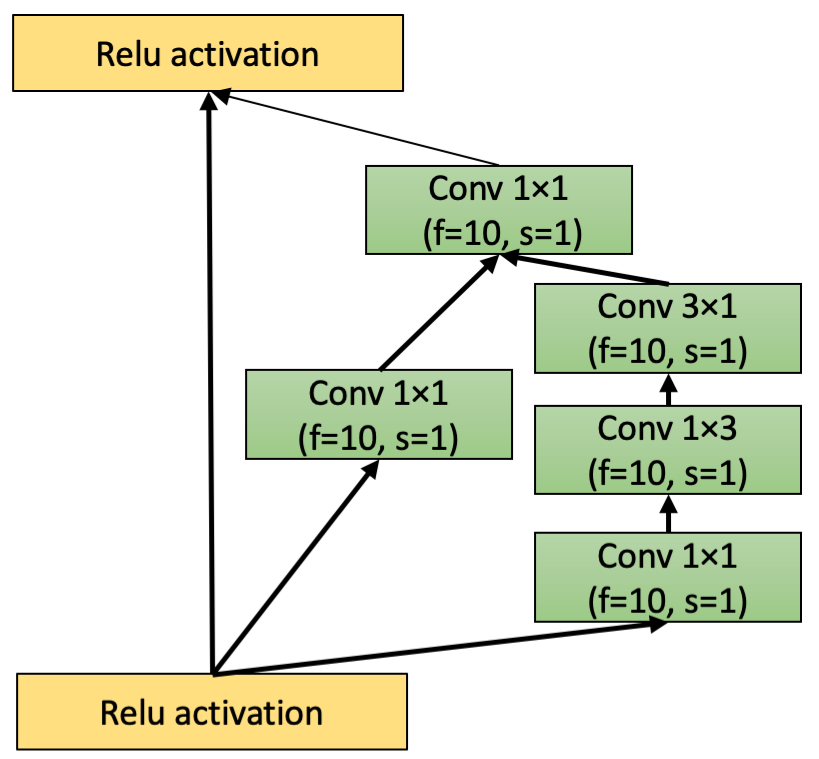}
    \captionof{figure}{The schema for Inception-ResNet-C, adapted from \cite{Inception_ResNet}.}
    \label{fig:C}
\end{figure}

\begin{figure}[hbt!]
    \centering
    \includegraphics[width=5cm]{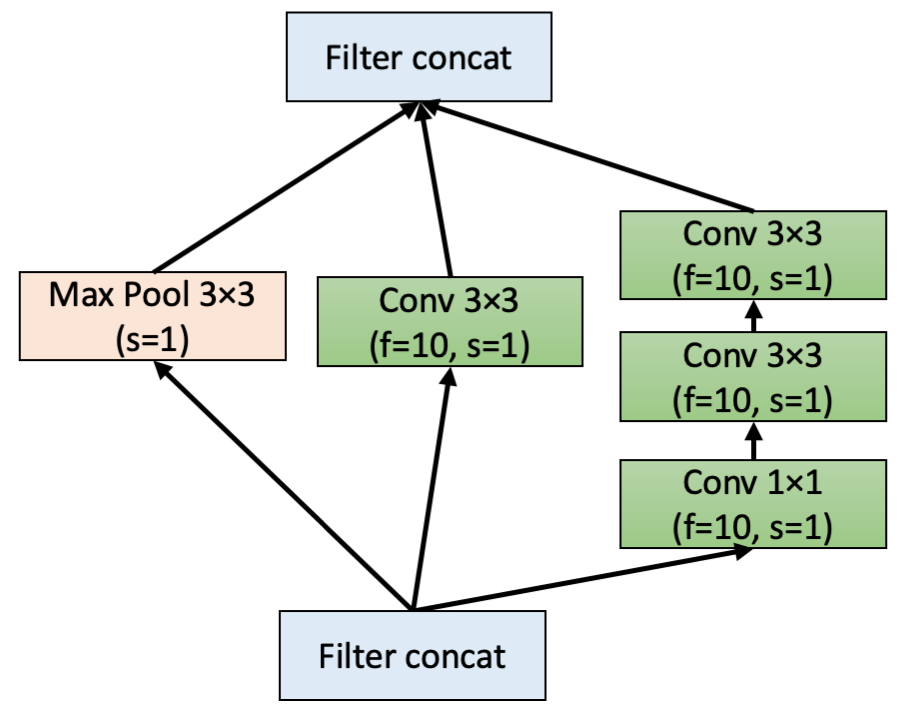}
    \captionof{figure}{The schema for Reduction-A, adapted from \cite{Inception_ResNet}.}
    \label{fig:reductions}
\end{figure}

Moreover, Inception-ResNet-C block (see Fig. \ref{fig:C}) is the equivalent to the Inception module without the $5\times5$ convolution layer. Note that the Max Pooling layer is again replaced by the ``shortcut connection'', and the $3\times3$ convolution is \textit{factorized} by $1\times3$ and $3\times1$ convolution layers. The module Reduction-A (see Fig. \ref{fig:reductions}) shrinks the number of parameters thanks a $3 \times 3$ Max Pooling layer.

Due to its deepness, the resulting Mini Inception-Resnet architecture is much more flexible than the one presented in \cite{Rome}. As we have discussed previously, increasing the number of layers might be counterproductive and would  drastically increase the computational complexity of the network. Nonetheless, these two concerns are solved with the incorporation of ``shortcut connections'', which allows the input not to be forgotten, and a  \textit{factorized} grid-like architecture that alleviates the computational complexity of the neural network.

\section{Data}

%

The Advanced LIGO-Virgo detector network collected data for three runs, during which more than 60 possible GW events have been identified \cite{Abbott:2020niy, LIGOScientific:2018mvr}. Almost all of them, if confirmed, are associated to the coalescence of binary systems with the peculiar chirp shape of the signal. This feature is used to extract the signal from the detector noise adopting a matched filter approach. In the case of CCSN data must be selected and processes using different algorithms. To enhance the detection probability and reduce the false alarm rate, the proposed method implies to select data in a time window around trigger times given by the neutrino observatories and take advantage of common GW features predicted by CCSN numerical simulations.

To assess the robustness of our method, we selected data from the second observing run of the Advanced GW detectors, without relaying on any neutrino information. In particular, we chosen a stretch of data taken during  August 2017, when Virgo joined the run \cite{Abbott:2019ebz}. The period includes about 15 days of coincidence time among the three detectors. 
About 2 years of time-shifts data have been used to construct a set of glitches to train and test the neural network. 

To build images for our neural network algorithm we use the internal features of cWB algorithm\footnote{cWB home page, \url{https://gwburst.gitlab.io/}; \\ public repositories, \url{https://gitlab.com/gwburst/public}\\ documentation, \url{https://gwburst.gitlab.io/documentation/latest/html/index.html}.}.
cWB is used by the LIGO and Virgo collaborations for the GWs search that uses minimal assumptions of the expected signal. It measures the energy excesses over the detector noise in the time-frequency domain and combines these excesses coherently among the various detectors of the network \cite{Abbott:2016ezn, Abbott:2019prv}. It is used in both low latency mode \cite{LIGOScientific:2019gag} and in triggered searches for CCSN \cite{SNTargeted2016,Abbott:2019pxc}.
In our work we used this software tool to compute the Wavelet transform, on the base of which the images of 256x64 pixels are build covering the frequency band from 0 o 2048 Hz and a time range of 2s.
Because the gravitational-wave signal is embedded in noise and difficult to extract, in the previous work \cite{Rome} a technique to visually enhance the coincidences among all the interferometers of the network has been developed. The method consists in using  primary colours for the spectrograms of each detector: red (R) for LIGO-Hanford, green (G) for LIGO - Livingston and blue (B) for Virgo.

The main results in this work present some differences with 
respect to \cite{Rome}: we incorporate the information of the source distance; we cover a larger parameter space with our phenomenological waveforms; we consider real data from the second scientific run, we do not anymore build images based on cWB selective information. The idea is to use the neural network as a follow up of multimessenger information.  
We define the starting time of each image every $2$~s, echoing the choice done in \cite{Rome}.  The images containing the central time of injected signals are considered as \textit{event class}, instead the ones without signals are the \textit{noise class}.
The injected signal is expected to be about 600-700 ms in duration, drawn everywhere in the image, with a small probability to be between two consecutive images. Such images are used for the training in any case, therefore the network can recognise also partial signature of the event.

We first compare the new architecture of the neural network using, for the training and validation, the same data set of the previous work \cite{Rome} (section \ref{sec:Old}). 
To tune our CNN we train the algorithm with the new phenomenological templates injected in the real noise (section \ref{sec:training}). Finally we test the network with injections of phenomenological waveforms (section \ref{sec:blind}) and waveforms from CCSN numerical simulations (section \ref{sec:testSet}).

\begin{table*}
\caption{List of models of the test set used in the injections. $M_{\rm ZAMS}$ corresponds to the progenitor mass at zero-age in the main sequence (ZAMS).
Unless commented, all progenitors have solar metallicity, result in explosions and their GW signal do not show signatures of the standing-shock accretion
instability (SASI).}
\label{tab:SNModels}
\begin{tabular}{cccc}
\hline
Model name & reference & $M_{\rm ZAMS}$& comments\\ 
\hline
s9 & \cite{Radice_2019} & $9 M_\odot$ & Low mass progenitor, low GW amplitude. \\
s25 &\cite{Radice_2019} & $25 M_\odot$ & Develops SASI.\\
s13 &\cite{Radice_2019} & $13 M_\odot$ & Non-exploding model.\\
s18 & \cite{Powell_2018} & $18 M_\odot$ & Higher GW amplitude. \\
he3.5 & \cite{Powell_2018} & - & Ultra-stripped progenitor ($3.5 M_\odot$ He core).\\
SFHx & \cite{Kuroda_2016} & $15 M_\odot$ & Non-exploding model. Develops SASI.\\
mesa20 & \cite{O_Connor_2018} & $20 M_\odot$& \\
mesa20\_pert & \cite{O_Connor_2018} & $20 M_\odot$ & Same as mesa20, but including perturbations.\\
s11.2 & \cite{Andresen2017} & $11.2 M_\odot$ &\\
L15 & \cite{Muller2012} & $15 M_\odot$ & Simplified neutrino treatment.\\ 
\hline
\end{tabular}
\end{table*}

\subsection{Previous set}
\label{sec:Old}

In \cite{Rome}, phenomenological supernova signals  were injected in Gaussian Noise simulating the final expected sensitivity of Advanced LIGO and Virgo detectors. Signals were injected at fixed network SNR, and did not included any information about source distance. This set was constructed using the information given by cWB algorithm and, unlike in the following data sets, only using events passing the first stage of cWB analysis.  This set contains about $10000$ images with signals for 11 different SNR ranging from $8$ to $40$ and the same amount with only noise, $75\%$ of the signals are used to train
the network and $25\%$ for validation. 

\subsection{Training set}
\label{sec:training}
The training set for CCSN signals has been constructed injecting waveforms at fixed distances: 0.2, 0.4, 1, 2 and 3 kpc. For this purpose, we have used the waveform template bank described in section~\ref{sec:waveforms} injecting, for each distance, of the order of  $70000$ waveforms, with random sky localization. $75\%$ of the set is used in the actual training while the remaining $25\%$ is used for validation.

\subsection{Blind Set}
\label{sec:blind}
In the blind set we injected a new ensemble of about 260000 simulated signals, generated by the phenomenological templates described in section~\ref{sec:waveforms}. In this case distance is chosen in a uniform distribution between 0.2 and 15 kpc, position in the sky are randomly chosen. This set is used to quantify the detection efficiency and to test the network. It is not involved in the training or validation procedure.

\subsection{Test set}
\label{sec:testSet}

For the final test we perform injections using CCSN waveforms from numerical simulations found in the literature. In particular we focus on 3D simulations of non-rotating progenitors representative of the neutrino driven mechanism. The selection {\it test set}, hereafter, see Table~\ref{tab:SNModels}, is performed based on the realism of the computed simulations in terms of neutrino transport and equation of state  and on the completeness of the GW signal\footnote{Some of the models in the literature
 compute less than $100$~ms after bounce or have a poor sampling rate.}. The selection includes models with a variety of mass progenitors and features
in the GW spectrum, and coincides with the choice for ongoing SN searches by the LIGO-Virgo-KAGRA collaboration. Except for model L15, none of the models coincide with the models
selected for the waveform calibration set used in Section~\ref{sec:waveforms}. With this choice the injected waveforms are in practice completely uncorrelated to any information 
we have used to train the CNN network.
The procedure is similar to the one used for the blind set of the previous test: we injected about 65000 waveforms uniformly in distance and sky directions, from 100 pc to 15 kpc.

\section{Training Methodology}


In this section we describe how we convert training images into categorical data for the identification of CCSN signatures in Gaussian and real noise, to solve our multiclass image classification task.

As in \cite{Rome}, we train the network using {\it{curriculum learning}}, where we start training with the easiest data sets, and then gradually the task difficulty is increased. We note that, although our new template bank is constructed using a series of fixed distances, the SNR follows, for each of these distances, a statistical distribution resulting from the random process used to generate the waveforms (see Sect.~\ref{sec:waveforms}). In practice, instead of using the distance, we define  data  as a set of templates that have SNRs in a fixed range. In this way, the difficulty of the data sets increases with decreasing SNR.
The data sets are balanced, so that $50\%$ of the templates belong to the event class and $50\%$ to the noise class. Because the present network is much larger then that in \cite{Rome} where we had balanced training and validation sets, here we use $75\%$ of the data for training and $25\%$ for testing.

\begin{table}[h!]
\caption{Confusion matrix for event and noise class} \label{tab:confusion}
\begin{tabular}{cc|c|c|c}

\cline{3-4}
                                                                                                          &                & \multicolumn{2}{c|}{\textbf{Actual class}}                                                                                      &  \\ \cline{3-4}
                                                                                                          &                & \textit{Event}                                                 & \textit{Noise}                                                 &  \\ \cline{1-4}
\multicolumn{1}{|c|}{\multirow{2}{*}{\textbf{\begin{tabular}[c]{@{}c@{}}Predicted\\ class\end{tabular}}}} & \textit{Event} & \begin{tabular}[c]{@{}c@{}}True\\  positive (TP)\end{tabular}  & \begin{tabular}[c]{@{}c@{}}False \\ positive (FP)\end{tabular} &  \\ \cline{2-4}
\multicolumn{1}{|c|}{}                                                                                    & \textit{Noise} & \begin{tabular}[c]{@{}c@{}}False \\ negative (FN)\end{tabular} & \begin{tabular}[c]{@{}c@{}}True\\  negative (TN)\end{tabular}  &  \\ \cline{1-4}
\end{tabular}
\end{table}

In the previous paper we measured the performance of the neural network in terms of the efficiency $\eta_{CNN}$ and the false alarm rate $FAR_{CNN}$, which are equivalent to the true positive rate and the false discovery rate, respectively. Here we will redefine these variables in terms of the confusion matrix (see Table \ref{tab:confusion}), but the definitions are completely equivalent.

\begin{equation}
    \eta_{CNN} = \frac{\textrm{correctly classified signals}}{\textrm{all the signals at CNN input}} 
    = \frac{TP}{TP+FN}
\end{equation}

\begin{equation}\label{eq:FAR}
    FAR_{CNN} = \frac{\textrm{misclassified noise}}{\textrm{all classified events}} = \frac{FP}{FP+TP}
\end{equation}

In this research we also measure the performance of our network with the \textit{receiver operating characteristic} curve (ROC curve), which is created by plotting the true positive rate (TPR) agains the false positive rate (FPR). Note that the definition of $\eta_{CNN}$ coincides with TPR, but FPR is defined as: 

\begin{equation}
    FPR  = \frac{FP}{FP+TN}
\end{equation}

 \section{Results}
 \subsection{Waveform injection in Gaussian noise: comparison with previous results}

In this subsection we will describe the experiments performed with injections in Gaussian noise. To train and validate the network, we use the data set  described in section \ref{sec:Old}, composed of waveforms ranging in the interval SNR=$[8, 40]$.  This choice allows for a direct comparison with the results in \cite{Rome} and it helps to improve the present software architecture.

To improve the performance of \cite{Rome} it is necessary to minimize $FAR_{CNN}$ while maximizing $\eta_{CNN}$. Therefore, from Eq. \ref{eq:FAR} we wish to minimize FP instead of FN, i.e. we need to penalize the algorithm when it classifies noise class as event class. To be able to penalize the algorithm we implement \textit{weighted binary cross-entropy}, where we assign weight $w$ to the noise class and weight 1 to the event class. We vary this parameter between $w=[1.0, 3.5]$, where $w=1$ would be equivalent to a normal \textit{binary cross-entropy} and $w=3$ would mean that it is 3 times more important to correctly classify the noise class rather than the event class.

\begin{figure}[h]
    \centering
    \includegraphics[width=8cm]{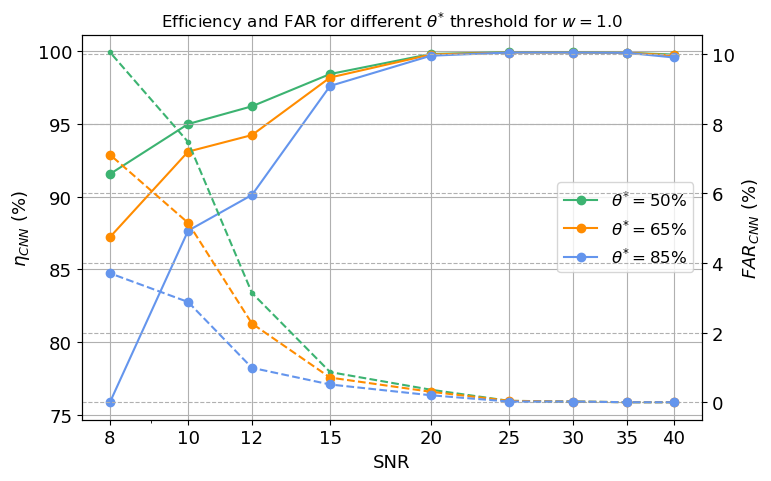}
    \captionof{figure}{$\eta_{CNN}$  (solid lines) and $FAR_{CNN}$ (dashed lines) for different SNRs  computed during the validation process for $w=1.0$ and different $\theta^\ast$ thresholds.}
    \label{fig:gaussian1}
\end{figure}

\begin{figure}[h]
    \centering
    \includegraphics[width=8cm]{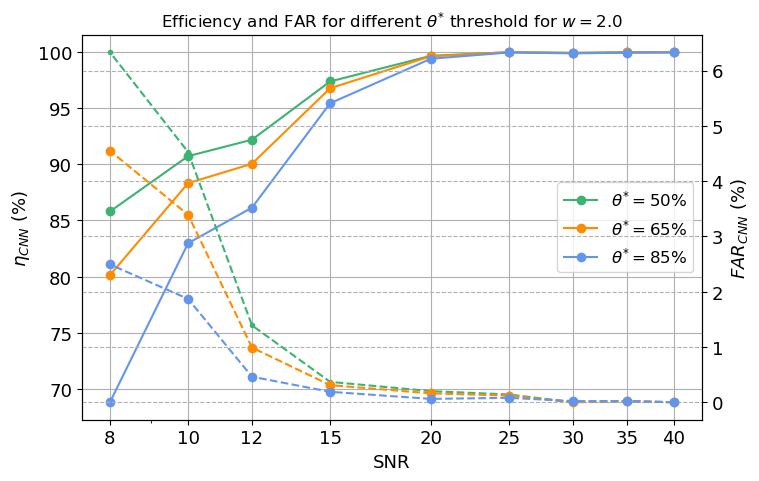}
    \captionof{figure}{$\eta_{CNN}$  (solid lines) and $FAR_{CNN}$ (dashed lines) for different SNRs  computed during the validation process for $w=2.0$ and different $\theta^\ast$ thresholds.}
    \label{fig:gaussian2}
\end{figure}

Moreover, the algorithm returns the probability $\theta$ that a certain template belongs to the event class. We want this probability to be high without dramatically decreasing $\eta_{CNN}$. Therefore, we define the decision threshold $\theta^\ast$ in range $ [50\%, 85\%]$; when a given probability exceeds this value, we will classify the template as an event, otherwise, it is classified as noise. 
Therefore, we perform different experiments to tune $w$ and $\theta$.
In figures \ref{fig:gaussian1} and \ref{fig:gaussian2}, we obtain $\eta_{CNN}$ and $FAR_{CNN}$ for $w=\{1,2\}$ and $\theta=\{50\%,65\%, 85\%\}$.

In Fig. \ref{fig:gaussian1}, we report the high performance of low $\theta$ in terms of $\eta_{CNN}$, paying the prize in even relative high $FAR_{CNN}$. The opposite behaviour occurs for high $\theta$. To be able to improve the probability distribution $\theta$, we will penalize the loss function with $w=2.0$. This means that the impact of correctly classifying noise templates is twice higher than correctly classifying event templates, as we show in Fig. \ref{fig:gaussian2} where the $FAR_{CNN}$ is minimized with respect to Fig. \ref{fig:gaussian1} with some cost in $\eta_{CNN}$.

Notice that $w$ will penalize the learning, so if the network is learning correctly the results would be enhanced, but it will lead to poor results otherwise.
This is  evident when we compare  the results shown in the figures \ref{fig:gaussian1} and \ref{fig:gaussian2}:  if we increase $w$ we  have less performance in terms of $\eta_{CNN}$, with little gains in $FAR_{CNN}$.
\\ 
To  have a clearer comparison between Fig. \ref{fig:gaussian1}, \ref{fig:gaussian2} and the results from the previous paper \cite{Rome}, we plot the validation results of Mini Inception Resnet for $w=\{1,2\}$ in Fig. \ref{fig:comparison}.

\begin{center}
    \centering
    \includegraphics[width=8cm]{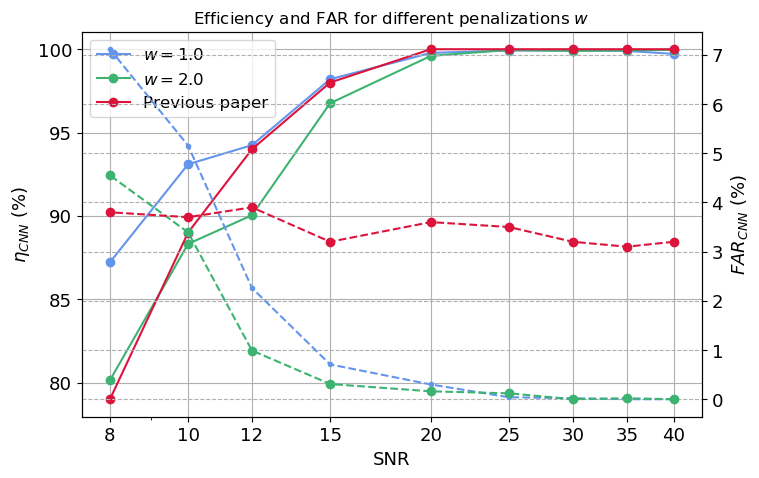}
    \captionof{figure}{$\eta_{CNN}$  (solid lines) and $FAR_{CNN}$ (dashed lines) as functions of SNR computed during the validation process of $w=\{1,2\}$, with $\theta^\ast =65 \%$,  and \cite{Rome}, where $\theta^\ast =50 \%$.}
    \label{fig:comparison}
\end{center}

Since we want to obtain a trade-off between $\eta_{CNN}$ and $FAR_{CNN}$, we settle $w=2.0$ and $\theta^\ast=65\%$.

The main improvement of our network with respect to \cite{Rome} is the minimization of $FAR_{CNN}$ towards $\sim 0\%$ for SNR in range $[15,20]$, while maintaining the same $\eta_{CNN}$. We note also that the poor performance at low SNR is due to the fact that this architecture is susceptible to the strong presence of Gaussian white noise, as it is pointed out in \cite{degradation}. Hence, the role of the decision threshold $\theta^\ast=65
\%$ is two-fold. On one hand, with this decision threshold we obtain $\max{(FAR_{CNN})} \approx 4 \% $  for low SNR which is the upper limit obtained by the previous paper \cite{Rome}. On the other hand, $\theta^\ast=65
\%$ provides us with a fair trade-off between $\eta_{CNN}$ and $FAR_{CNN}$ as we have discussed before. 

In terms of speed performance, in a GPU \textit{Nvidia} Quadro P5000 it takes 1h 18 min to train, validate and test  Mini Inception Resnet for this particular data set with 5 epochs for each SNR. A great part of this time is employed in training the neural network, so with bigger data sets the computational time will increase. Nonetheless, once the network is trained, the prediction is performed within minutes.

\subsection{Waveform injections in real detector noise: training and validation}

In this section we describe the experiments performed using the {\it training set} (section \ref{sec:training}).
This set contains injected phenomenological signals in real noise in the interval SNR=$[1, 232]$. As before, for each data set at a given SNR we calculate $FAR_{CNN}$ and $\eta_{CNN}$ during the validation. We also vary the penalization parameter $w \in \{1,2\}$ and
as in the previous section we choose $w=2$ and the decision threshold $\theta^\ast=65\%$.

For the network to learn correctly the input, it is crucial to perform a smooth ``curriculum learning''. Due to the difficulty of the data set, we separate the event templates into bins of size N and noise templates are packed accordingly. We performed the training for different $N$ but a better trade-off between $\eta_{CNN}$ and $FAR_{CNN}$ was observed for N=30.000, which provided a smoother transition between SNR bins. Therefore, in Fig. \ref{fig:real} we show the results of the validation process having fixed $N=30.000$, $\theta^\ast=65\%$ and $w=2$.

\begin{center}
    \centering
    \includegraphics[width=9cm]{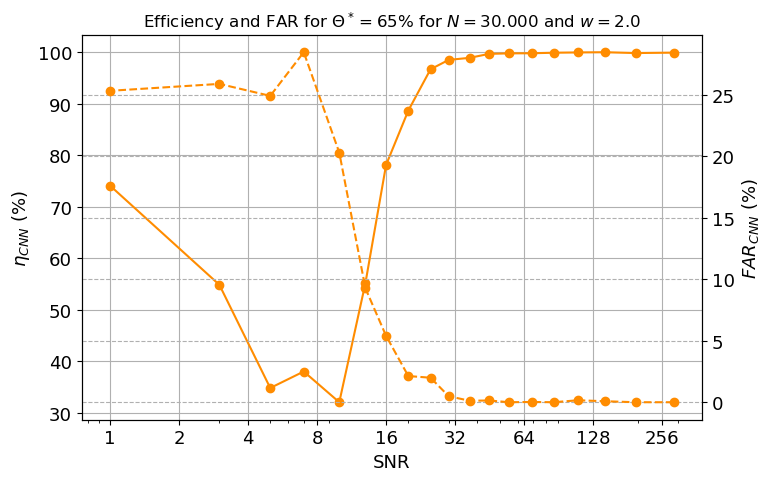}
    \captionof{figure}{$\eta_{CNN}$ (solid line) and $FAR_{CNN}$ (dashed line) for SNR in range $[1,232]$, for $w=2.0$ and $\theta^\ast=65\%$. These results are obtained from validating on 25$\%$ of the data that we have not trained on .}
    \label{fig:real}
\end{center}

In Fig. \ref{fig:real} we  note that $\eta_{CNN}$ is around $98\%$ above $SNR=32$ and below this value $\eta_{CNN}$ starts  decreasing. Instead, $FAR_{CNN}$ is around $0\%$ but increases for SNR values below 20. For lower SNR values of the network the method tends to show more and more an erratic behaviour 
that we foresee due to the statistical structure of the real noise.

This procedure is rather fast. In terms  of  speed  performance, in a GPU \textit{Nvidia} Quadro P5000 {it takes 2h 21 min for Mini  Inception-Resnet  to train and validate for this particular data set, but only 10 min to predict the \textit{blind set} and \textit{test set}. 
The time increase in the training phase is due to the fact that now we set the number of epochs to 10 instead of 5  to  guarantee  a  better  convergence of the network’s trainable parameters.}

\subsection{Waveform injections in real detector noise: final results}


In this section we present the results obtained when we used the network trained and optimized in the previous section on the data 
of the {\it blind set} (section \ref{sec:blind}) and the {\it test set} (section \ref{sec:testSet}). The network has not been trained by any of the images of these two sets so they can be used for the final test of the performance of the network. The signals injected in the {\it blind set} correspond to waveforms generated by the same procedure used to generate the training set, while the injections in the {\it test set} correspond to realistic CCSN waveforms. 

In Fig. \ref{fig:hist}, we report the histogram of the injections in the real noise. Such plot shows the robustness of the decision threshold $\theta^\ast=65\%$ even in the case of real detector noise. 

\begin{center}
    \centering
    \includegraphics[width=9cm]{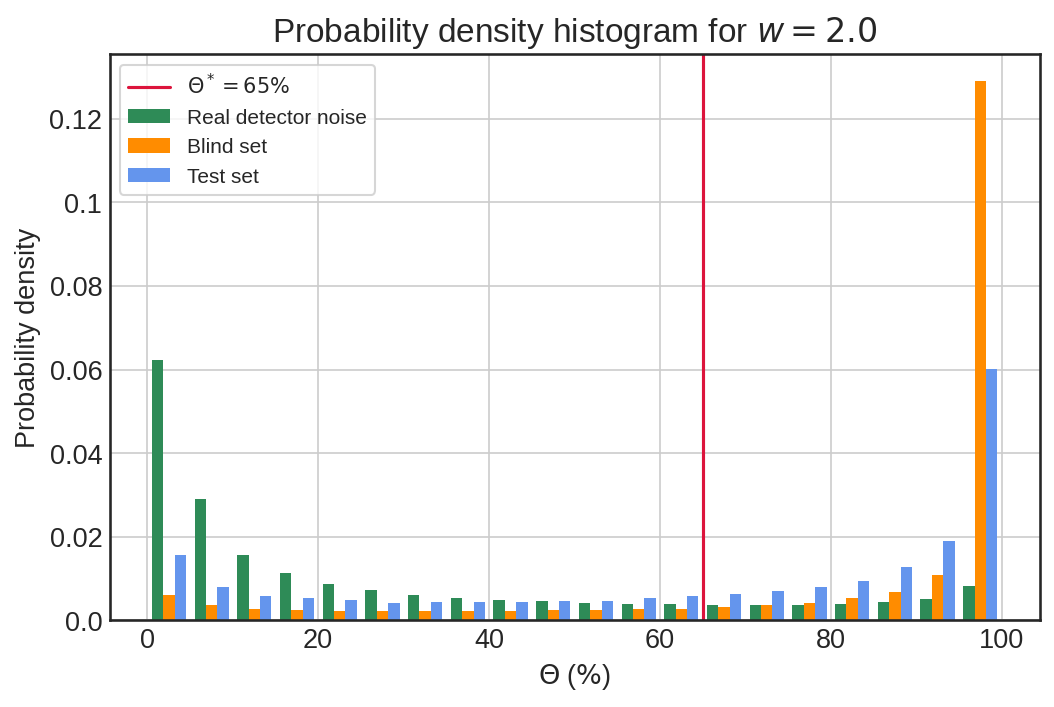}
    \captionof{figure}{Histogram of real detector noise and injections in real time as function of the probabilities predicted by Mini Inception Resnet. The vertical line represents the chosen decision threshold $\theta^\ast =65 \%$. Given the counts of the $i$th bin $c_i$ and its width $b_i$, we define the probability density  as $c_i/(\sum_i^N c_i \times b_i)$, where $N$ is the total number of bins of the histogram. }
    \label{fig:hist}
\end{center}

In Fig. \ref{fig:roc} we plot the Receiver Operating Characteristic (ROC) curve and we calculate the area under the curve (AUC). 

\begin{center}
    \centering
    \includegraphics[width=9cm]{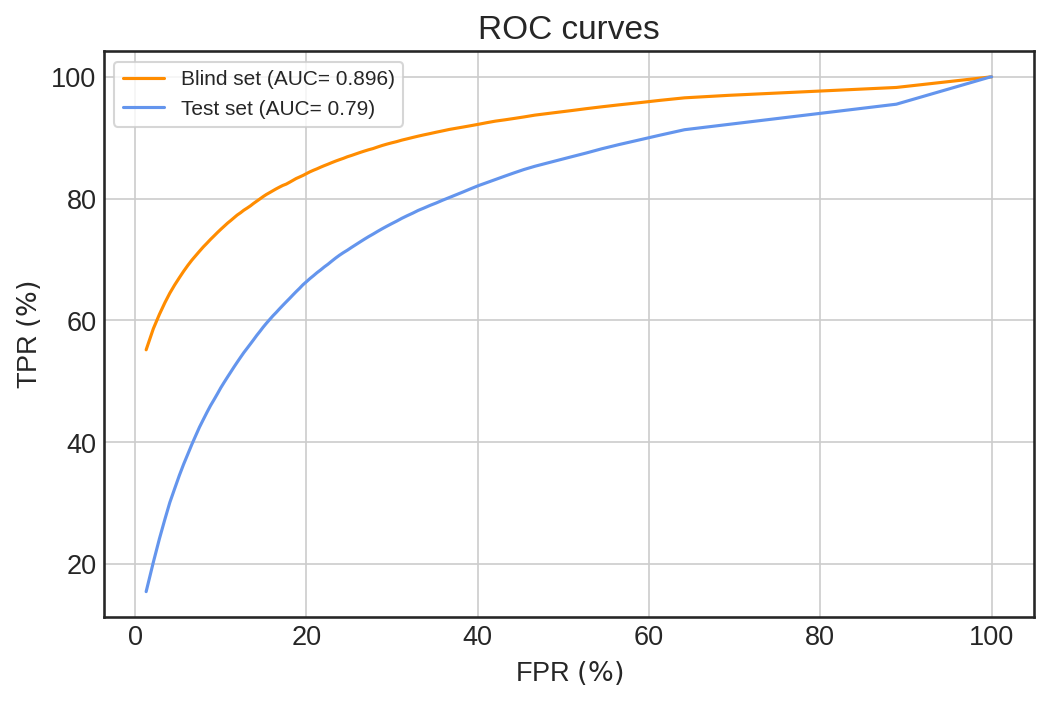}
    \captionof{figure}{Performance of our neural network for the {\it blind set}  and the  {\it test set}  for $\{w, N, \theta^\ast \}=\{2, 30.000, 65\% \}$. AUC is presented in the legend of the plot.}
    \label{fig:roc}
\end{center}

We note the high performance of the  {\it test set} (AUC=0.79) compared  with that obtained for the {\it blind set} (AUC=0.90). Even if we only trained our network with phenomenological waveforms from the template bank described in section~\ref{sec:waveforms}, such waveforms mimic the behaviour of the test set described in \ref{sec:testSet}, which is the main reason behind such good results. 

Another interesting graph that shows the resemblance between the {\it blind set} and {\it test set} is Fig. \ref{fig:eff_dist}. Here we plot $\eta_{CNN}$ as a function of the distance. 

\begin{center}
    \centering
    \includegraphics[width=9cm]{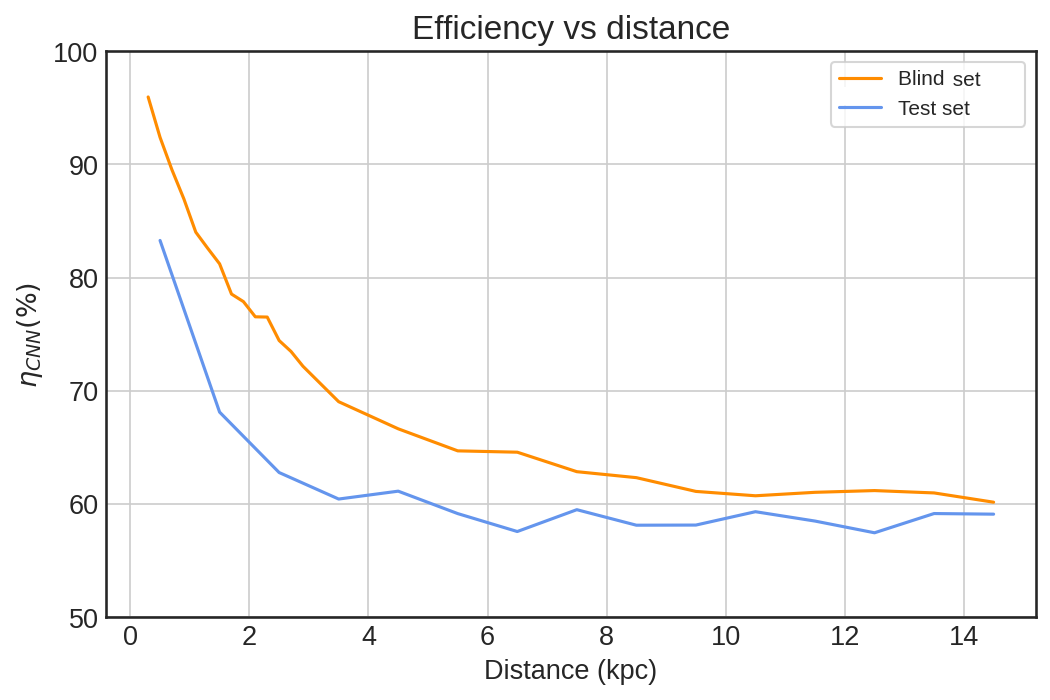}
    \captionof{figure}{$\eta_{CNN}$ as a function of the distance computed during the testing process for $\{w, N, \theta^\ast \}=\{2, 30.000, 65\% \}$.}
    \label{fig:eff_dist}
\end{center}

As we can see, at short distances there is a difference in efficiency between {\it blind set} and {\it test set} of $\approx 10 \%$, but when we increase the distance, they seem to reach a lower limit at $\eta_{CNN} \approx 60 \%$. In Fig. \ref{fig:eff_snr} we also plotted $\eta_{CNN}$ against SNR. 

\begin{center}
    \centering
    \includegraphics[width=9cm]{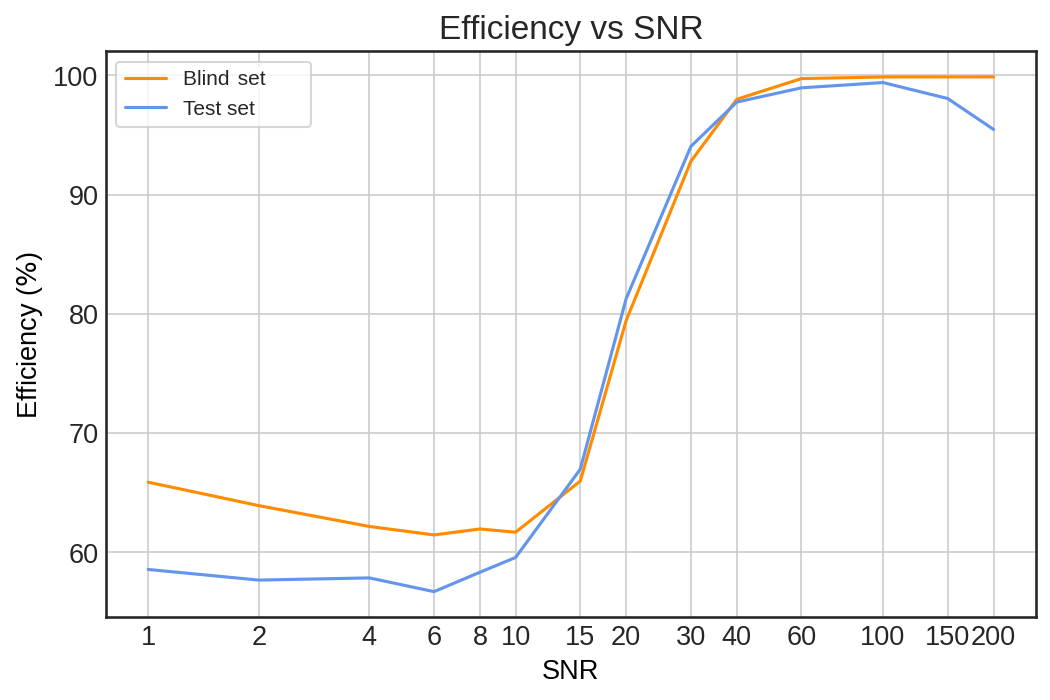}
    \captionof{figure}{$\eta_{CNN}$ as a function of SNR computed during the testing process for $\{w, N, \theta^\ast \}=\{2, 30.000, 65\% \}$. }
    \label{fig:eff_snr}
\end{center}

For low SNR, the difference in efficiency $\eta_{CNN}$ in the two case, {\it blind set} and {\it test set}, is around $10 \%$, while for SNR $>15$ we obtain the same efficiency. This final result assesses the robustness of  this method to detect CCSN signals embedded in the real detector  noise.  

\section{Discussion}


 The search of CCSN signals is carried by a software algorithm whose architecture trains $98.997$ parameters, significantly more than the $3.210$ parameters taken into account in \cite{Rome}. This implies an increases of the network complexity by a factor $30$. We trained the Mini Inception-Resnet using of about $26000$ images corresponding to spectrograms of phenomenological waveforms injected in real noise of the three detector network LIGO-Virgo during the second observation run, and similar number of images without signals. We used the curriculum learning with decreasing value of the SNR for the training. The significant differences with \cite{Rome} are: 
\begin{itemize} 
\item[-] the increase of the training images by a factor $\sim 10$, 
\item[-] the extended variability of the injected waveforms, to mimic the behaviour of the results from the CCSN numerical simulations, 
\item[-] the novel waveform parametrization for the frequency evolution,
\item[-] the use of real detector noise instead of Gaussian one,
\item[-] images are not anymore built by applying a SNR threshold by cWB.
\end{itemize}

First, to compare the efficiency of this new method with previous results, we run the Mini Inception-Resnet network with the same setup as in \cite{Rome}. The validation step shows that, with the appropriate choice
of parameters ($\theta^\ast=65\%$ and $w=2$) we  minimize the $FAR_{CNN}$ toward $\sim 0 ~\%$  almost maintaining the same efficiency $\eta_{CNN}$ for the range of SNR=$[15~,~20]$. 

Then, we tested this method  by injecting signals in the noise data of the LIGO-Virgo network taken during the second observation run.  
We have applied this analysis method  for detecting two classes of signals. The first one is  a blind set composed of  the same phenomenological templates having the  same analytical structure of those signals of the training set. The second one is based on 3D realistic numerical CCSNe simulations available in the literature. 
 
In the validation process, carried on using the dedicated  $25 ~\%$ of the training set where signals are uniformly distributed  in distance between 0.2 and 15 kpc, we obtain  about $80\%$ efficiency with a false alarm rate  of about $5\%$ for $SNR=16$, see figure \ref{fig:comparison}.


Applying the same method trained with phenomenological templates, to the case of realistic GW signals from 3D numerical simulations ({\it test set}) we still obtain a reliable performance. Overall, when compared to the case of the blind set, the efficiency at SNR$>15$, is very similar while at lower SNR we observe a reduction of less than $10\%$, see figure \ref{fig:eff_dist}.
The satisfying agreement is an indication that our phenomenological template
generator is  mimicking the main features observed in realistic CCSN and therefore, it supports the choice of this kind of
templates to train CNNs. The decrease in efficiency at low SNR could be an indication that some of the features of CCSN are not perfectly 
captured by the templates, for example the variability of the waveform amplitude for the duration of the signal (we consider that is in average
constant) or the presence of low frequency components associated to SASI. Future work could incorporate these two features to improve the performance of the search method. 

One of the advantages of the newly developed phenomenological templates is that they contain information about the distance to the source, which
allows us to study the performance of the blind set with respect to the distance and to compare directly with the results of the test set. 
With respect to the distance, the efficiency shows in general a quick drop at $\sim 2$ kpc followed by a gentle decline, falling down to 
about $60\%$ at $15$ kpc. This contrasts very much with the behaviour with respect to SNR that shows a step decline at SNR$\sim 15$. The reason 
for this difference is that, at a given distance, there is some variability in the amplitude of the possible waveforms, which tends to smooth 
out the results over a range of distances. We  expect that at larger distances the efficiency will keep decreasing towards zero, but we did not
see this effect within the limited set of distances used in this work. The performance with the {\it realistic} test set is somewhat worse than with 
the blind set, but the difference in efficiency is never larger than $10\%$, in agreement with the results obtained as a function of SNR.


We note that this results have been obtained using realistic waveforms from 3D models, which are in general about a factor $5$
weaker than those of 2D simulations. It is also important to notice that we have used real O2 noise, so the results are expected to be better for the current detector configuration, which recently ended O3, and will improve further once the final sensitivity of LIGO, Virgo and KAGRA detectors will be  achieved.
These two factors make it difficult to compare our results with those obtained in other papers using injections based on 2D simulations, simulated Gaussian noise and/or ultimate detector sensitivity \cite{Chan:2019,Cavaglia:2020,Iess:2020}.

In fact, for the case of neutrino driven explosions in \cite{Chan:2019} they use a set of waveforms from $55$ numerical simulations (mixed 2D and 3D) to perform about $10^5$ injections with random orientations in the sky in the range $0.2$-$200$~kpc. Using a LIGO-Virgo-KAGRA network with optimal sensitivity, they obtain an efficiency of $50\%$ at $4$~kpc with false alarm probability of $0.1\%$.  These results are similar to  our work, however it is difficult a closer comparison since they are using an interferometer network with ultimate sensitivity.

The work of \cite{Cavaglia:2020} focused in using Genetic Programming algorithms to improve the significance
of a single interferometer detection. For that purpose they trained the algorithm making injections of CCSN waveforms in real detector noise from the LIGO/Virgo first observing run (O1). For the case of neutrino driven explosions the algorithm is trained using waveforms from 2D and 3D CCSN simulations (8 in total) injected at different locations in the sky and distances in the range $1$-$7.5$~kpc (about $15000$ injections in total).
Similarly to \cite{Rome}, they employed cWB pipeline. For waveforms from 3D simulations (not the same as ours) they get an efficiency of $86\%$ at $3.16$~kpc with $12\%$ of false negatives. Again, the results are in the bulk of our numbers but it is difficult to compare, since they are using a network with lower sensitivity that ours and the injections that are comparable to ours amount only to 4 different signals.
Their results show that it is possible a detection with high significance ($3$-$\sigma$) for signals with an SNR as low as $10$. However, it should be noted that, in their case, the same waveforms were used for training and for testing.

Finally, \cite{Iess:2020} utilized a CNN trained using 5 waveforms from neutrino-driven CCSN 3D simulations injected in Gaussian noise considering the spectral sensitivity curve of Virgo during the third observing run (O3). Training was performed with about $25000$ random injections in the sky at distances between $0.01$ and $10$~kpc. To test the robustness of the method they also accounted for short duration detector noise transients, known as glitches, in simulated data.
When using different waveforms for training and testing, they obtain an efficiency of $\sim90\%$ of all triggers with a $\sim 10\%$ false alarms (all distances in the range). 
When using the same waveforms for testing and training they observe a drop in the efficiency, below $50\%$, for values of the SNR in the range $11-16$,
depending on the waveform.

Despite the differences with earlier works, overall our results seem consistent with other machine learning approaches. The drop of the efficiency at SNR$\sim 10-15$ is common for all algorithms (except for \cite{Chan:2019} that do not show this metric), which makes one wonder if there is some intrinsic limitation of machine learning algorithms that prevents to get closer to SNR$\sim 8$, typical value for optimal template-matching algorithms. It could also be possible that more complex architecture or training sets with different pixel resolutions might improve the efficiency of this method. These are aspects that we would like to explore in the future.

\section{Conclusion}
\label{sec:conclusion}
We developed a new machine learning algorithm to further improve the detectability of a GW signal from CCSN, following the path traced in \cite{Rome}.
Regarding the applicability of our method for the GW detection, we have considered a detection threshold, $\theta^\ast=65\%$, that results in
a FAR of about $5\%$ at SNR$\sim 15$  (or a FPR of $\sim 10\%$ at TPR$=50\%$). These values could be appropriate for an observation with 
high confidence of an event in coincidence with a neutrino signal. In those cases the neutrino signal is expected to be bounded within
$20$ s during the initial SNEWS alert \citep{SNEWS} and very likely well within $1$ s in the detailed analysis of high sensitivity neutrino detectors 
such as Super-K \cite{SuperK}. If the method were to be used in all-sky non-triggered searches, the range of values of FAR needed to make a
detection with high confidence could be achieved by using values of $\theta$ very close to $100\%$. The efficiency of the algorithm in this
regime is something that could be explored in future work.

These results are very promising for future detections of GWs from CCSN, because the network allows us to observe more than half of the events
within $15$ kpc.

This work has multiple possible extensions. At present the entire data processing is rather fast: the training and validation phase, performed in the real detector noise, is done in 2 hours and 21 minute using a GPU \textit{Nvidia} Quadro P5000, while  predicting the test set takes 3 ms for each 2 s long image. 
Given that we take advantage of  the Keras/TensorFlow framework, widely used within the 
machine learning community, it should be easy to increase the complexity of our current CNN or to incorporate the latest developments in 
machine learning algorithms, with a reasonable increase of the computational cost of the signal search. 
Furthermore, we could increase the number of classes to be able to detect other GW sources with the same architecture.
In the future, the new algorithm presented here should be compared under realistic conditions with the methods currently in use within the LIGO-Virgo collaboration
to evaluate the real advantages of the method. In particular, CNNs have the advantage that, once the training phase is performed, they have a very low computational cost, which could provide an advantage in the design of new low-latency detection pipelines for CCSN.

\vskip 2mm
\section{Acknowledgement}
The authors would like to thank M. Razzano for the critical reading of the manuscript and for his
constructive inputs.
This research has made use of data, software and/or web tools obtained from the Gravitational Wave Open Science Center (//https://www.gw-openscience.org/ /), a service of LIGO Laboratory, the LIGO Scientific Collaboration and the Virgo Collaboration. LIGO is funded by the U.S. National Science Foundation. Virgo is funded, through the European Gravitational Observatory (EGO), by the French Centre National de Recherche Scientifique (CNRS), the Italian Istituto Nazionale della Fisica Nucleare (INFN) and the Dutch Nikhef, with contributions by institutions from Belgium, Germany, Greece, Hungary, Ireland, Japan, Monaco, Poland, Portugal, Spain.
PCD acknowledges the support from the grants PGC2018-095984-B-I00, PROMETEU/2019/071 and the {\it Ramon y Cajal} funding (RYC-2015-19074) supporting his research.
In addition, IDP and FR acknowledge the support from the Amaldi Research Center funded by the MIUR program "Dipartimento di Eccellenza" (CUP:B81I18001170001) and the Sapienza School for Advanced Studies (SSAS).
\bibliographystyle{unsrt}
\bibliography{references}



\end{document}